\def\year{2022}\relax
\documentclass[letterpaper]{article} 
\usepackage{aaai22}  
\usepackage{times}  
\usepackage{helvet}  
\usepackage{courier}  
\usepackage[hyphens]{url}  
\usepackage{graphicx} 
\usepackage{natbib}  
\usepackage{caption} 
\DeclareCaptionStyle{ruled}{labelfont=normalfont,labelsep=colon,strut=off} 
\usepackage{algorithm}
\usepackage{algorithmic}
\usepackage{amsmath}
\usepackage{multirow}
\DeclareMathOperator*{\argmin}{arg\,min}
\usepackage[flushleft]{threeparttable}
\usepackage{newfloat}
\usepackage{listings}
\floatstyle{ruled}
\newfloat{listing}{tb}{lst}{}
\floatname{listing}{Listing}
\usepackage{color}

\title{Interpretable Knowledge Tracing: \\ Simple and Efficient Student Modeling with Causal Relations}

\title{Interpretable Knowledge Tracing: \\ Simple and Efficient Student Modeling with Causal Relations}
\author {
    Sein Minn\textsuperscript{\rm 1}
    Jill-Jênn Vie\textsuperscript{\rm 2}
    Koh Takeuchi\textsuperscript{\rm 3}
    Hisashi Kashima\textsuperscript{\rm 3}
    Feida Zhu\textsuperscript{\rm 4}
}
\affiliations {
    \textsuperscript{\rm 1}  Univ. Lille, Inria, CNRS, Centrale Lille, UMR 9189
- CRIStAL, F-59000 Lille, France\\
    \textsuperscript{\rm 2} Université Paris-Saclay, Inria, CEA, 91120 Palaiseau, France\\
    \textsuperscript{\rm 3} Kyoto University, Japan\\
    \textsuperscript{\rm 4} Singapore Management University, Singapore\\
    \{sein.minn, jill-jenn.vie\}@inria.fr, \{takeuchi, kashima\}@i.kyoto-u.ac.jp, fdzhu@smu.edu.sg
}

\usepackage{bibentry}

\begin{document}

\maketitle

\begin{abstract}
Intelligent Tutoring Systems have become critically important in future learning environments. Knowledge Tracing (KT) is a crucial part of that system. It is about inferring the skill mastery of students and predicting their performance to adjust the curriculum accordingly. Deep Learning-based KT models have shown significant predictive performance compared with traditional models. However, it is difficult to extract psychologically meaningful explanations from the tens of thousands of parameters in neural networks, that would relate to cognitive theory. There are several ways to achieve high accuracy in student performance prediction but diagnostic and prognostic reasoning are more critical in learning sciences. Since KT problem has few observable features (problem ID and student's correctness at each practice), we extract meaningful latent features from students' response data by using machine learning and data mining techniques. In this work, we present Interpretable Knowledge Tracing (IKT), a simple model that relies on three meaningful latent features: individual \textit{skill mastery},  \textit{ability profile} (learning transfer across skills) and \textit{problem difficulty}. IKT's prediction of future student performance is made using a Tree-Augmented Naive Bayes Classifier (TAN), therefore its predictions are easier to explain than deep learning-based student models. IKT also shows better student performance prediction than deep learning-based student models without requiring a huge amount of parameters. We conduct ablation studies on each feature to examine their contribution to student performance prediction. Thus, IKT has great potential for providing adaptive and personalized instructions with causal reasoning in real-world educational systems.  
\end{abstract}

\section{Introduction}

Learning with a computer plays an essential role in today's education. Learning can be personalized with adaptive learning instructions to improve individual learning gains and enhance students' learning experience. A personalized adaptive learning environment is much more efficient than traditional learning environments like classroom learning~\cite{bloom19842}. Intelligent tutoring systems need to address the huge challenge of large-scale personalization for the process of real-world human learning. The most successful tutoring systems are utilized by tens or hundreds
of thousands of students a year with growing numbers~\cite{baker2016stupid}. To fulfill the requirements of personalization, we need an efficient method to assess the mastery state of students' skills (knowledge) empirically. 
Knowledge Tracing (KT) is used to  dynamically assess students' knowledge mastery state based on their past test outcomes. It can be used in the prediction of whether students may or may not answer the next problems correctly to adjust their personalized curriculum. 
AI techniques have found their way into the building of adaptive learning environments, in particular for the problem of knowledge tracing, for modeling students' conceptual or procedural knowledge from their observed performance on tasks~\cite{corbett1994knowledge}. The most current and well-known example is Deep Knowledge Tracing (DKT)~\cite{piech2015deep} with the use of Recurrent Neural Networks (RNNs).
Besides, it shows greater success in student performance prediction than early student models~\cite{piech2015deep,minn2019dynamic,minn2020bkt}. 

Bayesian Knowledge Tracing (BKT) is the earliest and well-known sequential approach to KT with psychologically meaningful parameters.  More specifically, BKT is a Hidden Markov Model (HMM), consisting of observed and latent variables that represent the student knowledge state on one specific skill (i.e., a skill can either be mastered by the student or not) and observed variables are assumed to
be binary (the student can answer the associated problem correctly or not)~\cite{corbett1994knowledge}. Several extensions of BKT have been introduced by using contextualized guessing and slipping parameters~\cite{d2008more}, estimated transition with the use of help features~\cite{baker2009state}, initial probability that the student knows the skill~\cite{pardos2010modeling}, item difficulty~\cite{pardos2011kt}, clusters of different student groups~\cite{pardos2012clustered}, student-specific parameters~\cite{yudelson2013individualized}. However, these extensions treat skills independently and are unable to detect learning transfer across skills.

Deep Knowledge Tracing (DKT) has obtained considerable attention thanks to its significant ability to model some of the learning transfer occurring across skills, which BKT cannot deal with.
However, because DKT feeds all of past students' interactions (binary values with associated skills) to RNNs, it cannot provide a psychological interpretation~\cite{piech2015deep} as the extensions of BKT do. 
Such psychological information is concealed in the hidden layer of RNNs with tens of thousands of parameters~\cite{khajah2016deep}.
Besides, several deep learning-based knowledge tracing models have been proposed in recent years: Deep Knowledge Tracing and Dynamic Student Classification~\cite{minn2018deep} enhances DKT with clusters of student ability profiles on skills at each time interval; Prerequisite-Driven Deep Knowledge Tracing~\cite{chen2018prerequisite} augments KT model by integrating prerequisite relations between skills;  Exercise-Enhanced Recurrent Neural Network with Attention mechanism~\cite{su2018exercise} computes a weighted combination of all previous knowledge states;  Sequential Key-Value Memory Networks~\cite{abdelrahman2019knowledge} is a hop-LSTM architecture that aggregates hidden knowledge states of similar problems into a new state;  Deep hierarchical knowledge tracing~\cite{wang2019deep} captures the relations between questions and skills to get problem representations; Graph-based Interaction Knowledge Tracing~\cite{yang2020gikt} utilizes graph convolutional networks to substantially incorporate question-skill correlations. 
Deep learning-based models show better predictive performance than the preceding approaches, in part because they are known to preserve past information in sequential data, such as student outcome traces. Nevertheless, those models are less likely to provide psychologically meaningful explanations for their inference than predicting the correctness of a given problem.

In this paper\footnote{This work is available at \url{https://github.com/simon-tan/IKT}}, we are trying to provide meaningful explanations through feature engineering and a simple probabilistic graphical model. Hence, we propose a novel student model called Interpretable Knowledge Tracing (IKT), by utilizing three meaningful features: individual  \textit{skill mastery}, \textit{ability profile} of students (learning transfer across skills), and  \textit{problem difficulty}. 
We first utilize conventional machine learning techniques, such as hidden Markov models and $k$-means clustering, to extract meaningful features, and then incorporate the extracted features using a Tree Augmented Naive Bayes classifier for inferring the correctness of a future problem. 
In contrast to the family of DKT, our IKT is a novel model to provide inference interpretation through a probabilistic graphical model with meaningful features while keeping high achieving prediction performance in the task of student performance prediction. We experimentally show that IKT outperforms well-known student models for performance prediction on several well-known knowledge tracing datasets. Additionally, we also conduct ablation studies to measure the contribution of each feature, by learning different tree structures.


\section{Background}\label{sec:back}
In successful learning environments like Cognitive Tutor and ASSISTments,  KT plays as a mechanism for tracing learners' knowledge~\cite{desmarais2012review}.


KT can be seen as a supervised sequential learning problem. The KT model is given student past interactions with the system that include: skills $S=(s_1,s_2,..,s_t) \in \{1, \ldots, M\}^t$ along with responses $R=(r_1,r_2,..,r_t) \in \{0, 1\}^t$  and predicts the probability of getting a correct answer for the next problem. It mainly depends on the mastery of the skill $s$ associated with problems  $P=(p_1,p_2,..,p_t)$. So we can define the probability of getting a correct answer as $p(r_{t}=1|s_{t},X)$ where $X=(x_1,x_2,..,x_{t-1})$ and $x_k= (s_k,r_k)$ is a tuple containing response $r_k$ to skill $s_k$ at time $k$. Then, we review here the best known KT modeling methods for estimating student's performance.

\begin{itemize}

\item {Item Response Theory (IRT).} In standardized tests, students' proficiency is assessed by one static latent variable~\cite{hambleton1991fundamentals}. IRT has a strong theoretical background both in terms of being grounded in psychometric measurement and relying on a sound mathematical framework. \citet{wilson2016back} proposed a Bayesian extension of IRT (BIRT) that shows competitive performance to DKT in terms of student performance prediction.

\item{Bayesian Knowledge Tracing (BKT)} is the earliest sequential approach to model a learner's changing knowledge state and is arguably the first model to relax the assumption on static knowledge states~\cite{corbett1994knowledge}. 

\item{Performance Factors Analysis (PFA)} was adapted from Learning Factor Analysis (LFA) ~\cite{cen2006learning} with sensitivity to the indicator of student learning performance. It allows conjunction by summing the contributions from all skills needed in a performance by relaxing the static knowledge assumption and modeling multiple skills simultaneously~\cite{pavlik2009performance}. 

\item{Deep Knowledge Tracing (DKT)} ~\cite{piech2015deep} uses a Long Short-Term Memory (LSTM) to represent the latent knowledge space of students along with the number of practices dynamically. This model compactly encodes the historical information from previous time steps by using the input, forget and output gates of a LSTM.

\item{Dynamic Key-Value Memory Networks (DKVMN)}  was proposed as an alternative to DKT that is inspired by the memory network architecture~\cite{zhang2017dynamic}.  It utilizes an external memory neural network module and uses two memory slots called key memory and value memory to encode the knowledge state of students. Assessments of knowledge state on particular skills are stored in memory slots and controlled by read and write operations through additional attention mechanisms. 

\item{ Deep Knowledge Tracing and Dynamic Student Classification  (DKT-DSC)}  was proposed to enhance DKT by utilizing dynamically evaluated student ability profiles at each time interval~\cite{minn2018deep}.  It applies $k$-means clustering to detect student ability profiles and takes them into account in student performance prediction.

\item {Context-Aware Attentive Knowledge Tracing (AKT-R)} was proposed by~\citet{ghosh2020context} and uses a monotonic version of the scaled dot-product attention mechanism for the encoding and retrieving of the knowledge state. It also applies Rasch model for learning the skill and problem embeddings. 

\end{itemize}

\section{Interpretable Knowledge Tracing}\label{sec:model}
When a student learns with an intelligent tutoring system (ITS), they practice a specific skill through answering several questions, and the ITS checks their mastery of skill according to whether they were able to provide correct answers.
However, even with a high level of mastery of the skill associated with some problems, the student may provide an incorrect answer to those problems. 
We tend to regard such situations as misunderstanding of the problem, or failure to utilize the related skill properly in a particular problem under a new circumstance. So we can assume that other factors, such as ability profile (learning transfer across skills) or difficulty of the occurring problem at the current timestamp, have a direct effect on the situation.



To adopt the aforementioned assumptions into ITS, we propose a feature extraction procedure consisting of three data mining techniques. 
Then, instead of feeding all data to neural networks in deep learning-based KT models, we propose a student model called Interpretable Knowledge Tracing (IKT) which can predict a student's future responses by relying on three meaningful latent features: individual \textit{skill mastery} of a student, \textit{ability profile} (across skills) and \textit{problem difficulty}.

\subsection{Interpretation through Feature Engineering}
Our procedure extracts three meaningful latent features from students' response data: \textit{skill mastery}, \textit{ability profile}, \textit{problem difficulty}. These features tell us about how much a student knows about the practicing skill, what kind of ability that student possesses, and how difficult the occurring problem is at each timestamp. Instead of taking students' past interaction sequence (binary values) and learning all information in a hidden state of deep learning-based KT model, the correctness of the problem is inferred by using these latent features as evidence at each timestamp.


\subsubsection{Knowledge tracing}\label{sec:skillmastery}
The formulation of skill mastery is inspired by the assessment of skill mastery (probability of learning a skill $s_t$) in Bayesian Knowledge Tracing (BKT), which is a well-known knowledge tracing model with psychologically meaningful parameters based on a Hidden Markov Model.
BKT infers mastery states, from ``not learned'' to ``learned'' and the probabilities above depend both on fixed parameters and the state at timestamp $t$. 

 
For a certain skill $s \in S$, BKT consists of four parameters representing probabilities:
\begin{itemize}
	\item$P(L_{0})$: the probability that a student masters the skill before attempting the first problem associated with $s$; 
	\item$P(T)$: the probability that a student, who does not currently master the skill, will master the skill after the next practice opportunity;
	\item$P(G)$: the probability that a student guesses the correct answer to a question despite not knowing the skill (\textit{guess});
	\item and $P(S)$: the probability that a student answers a question incorrectly despite knowing the skill (\textit{slip}).
\end{itemize}

We apply a brute-force search algorithm to fit BKT. We have $P(L_{0}), P(T), P(G), P(S)$ of each skill after fitting BKT. BKT is based on skill-specific modeling and can provide skill mastery of each skill according to the observed outcome $obs$ that is correct ($obs=1$) or incorrect ($obs=0$):

\begin{equation} \label{equ:corr}
P(L_{t}|1) = \frac{P(L_{t})(1-P(S))}{P(L_{t})(1-P(S))+(1-P(L_{t}))P(G)}
\end{equation}

\begin{equation} \label{equ:incorr}
P(L_{t}|0) = \frac{P(L_{t}) P(S)}{P(L_{t})P(S)+(1-P(L_{t}))(1-P(G))}
\end{equation}

\begin{equation} \label{equ:action}
P(L_{t+1}) = P(L_{t}|obs)+ (1-P(L_{t}|obs))P(T)
\end{equation}

By combining these equations, we can define the skill mastery as:
 \begin{equation} 
\textbf{skill mastery}(s_t)=\delta(P(L_{t}),s_t)
\end{equation}

\noindent
where $\delta(P(L_{t}),s_t)$ is a function that maps the skill mastery of particular skill $s_t$ at current timestamp in the whole student interaction. 

Note that \textit{skill mastery} is the probability of learning skill $s_t$ rather than the probability that a student applies the skill correctly in BKT. A BKT model is trained for each skill, and the inputs to each skill model are the binary responses of a student on that single skill. Other interleaved skills during the whole practice are ignored. Each skill model is independent, so there is no consideration of learning transfer across skills in this component.

\begin{figure}[t]
	\centering
	\includegraphics[trim={0mm 30mm 0mm 0mm}, clip, width=1\linewidth]{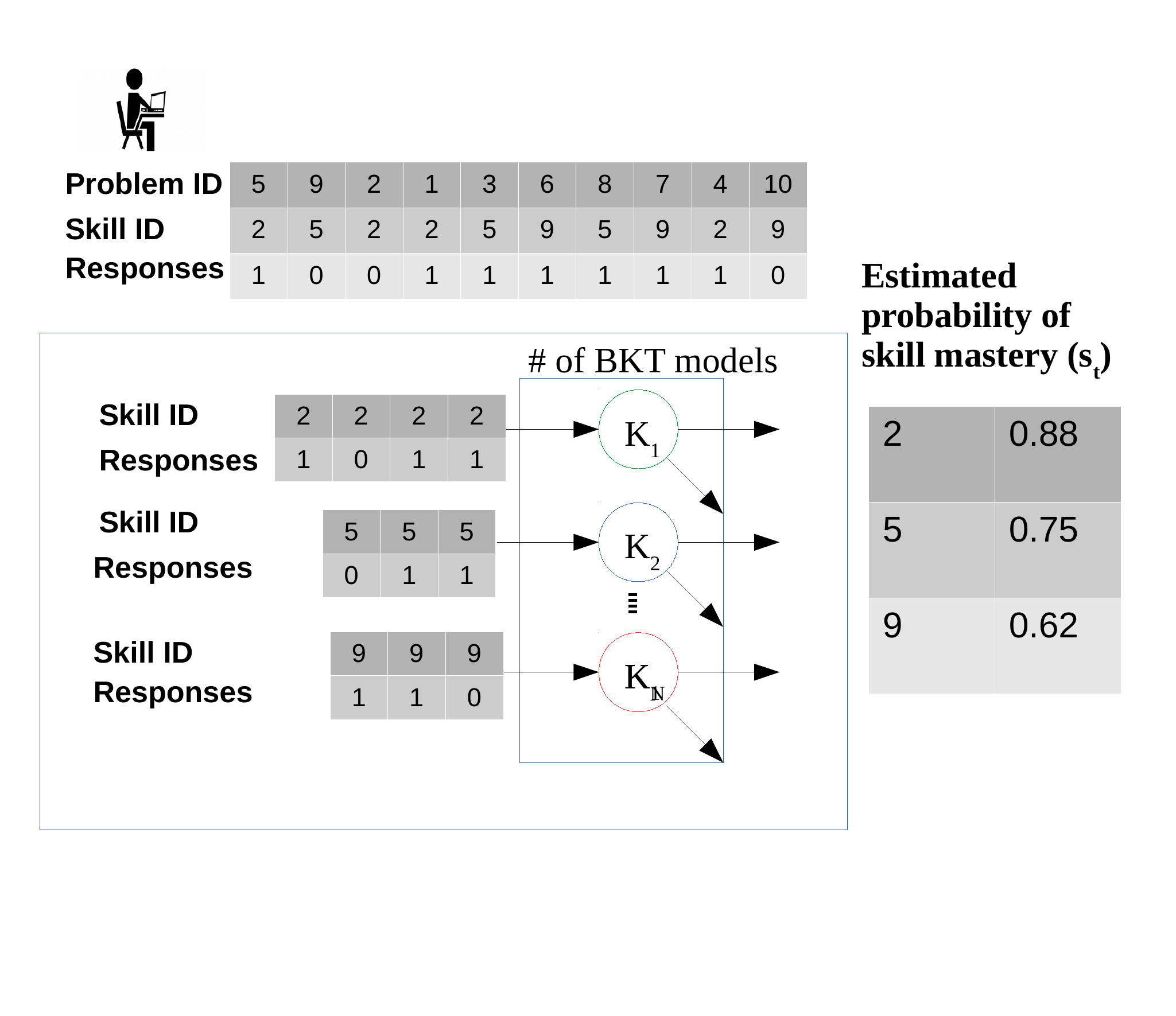}
	\caption{Skill mastery assessment for skills at each timestamp. Modeling of each skill is done independently and they do not interact with each other.}
	\label{fig:BKT2} 
\end{figure}

\subsubsection{Ability profiling}\label{sec:DTAB}

A strong limitation of BKT is that it treats each skill as an independent skill model, thus it can not discover or
leverage inter-skill similarity. When a student is assessed on a new skill, BKT considers the learning that they have gained from previous skills as irrelevant to the new skill.
Intuitively, learning transfer across skills is a naturally occurring phenomenon. General domain knowledge accumulates through practice and allows the student to solve new problems in the same domain with greater success, even if that problem involves new and specific skills. 
However, BKT and most KT models lack a mechanism to deal with the \textit{learning transfer} mechanism, and thus they cannot trace students' knowledge under transfer.
 
Learning transfer implies that students can transfer their acquired skills to new situations and across problems involving a different skill set.  ITS models such as Learning Factor Analysis (LFA)~\cite{cen2006learning} and Performance Factor Analysis (PFA)~\cite{pavlik2009performance} aim to capture this learning transfer phenomenon. 
They introduced a factor that represents the learning accumulated on all skills through practice and then utilized this factor as a predictor of success in further practice. 
These models have outperformed the standard BKT model without the skill transfer mechanism, and have shed new light on the importance to consider skill transfer.

When BKT performs the prediction task, it estimates the skill mastery of a student without considering learning transfer across skills.
This independent assumption makes BKT unable to evaluate any types of learning transfer that students should achieve at the current time interval in the long-term learning process. 
In DKT, an LSTM encodes the temporal information of student knowledge state with transfer learning in a single state vector and performs state-to-state transition globally. It is unable to assess the mastery of skills and the ability profile of the student to transfer learning. 

To detect the regular changes of learning transfer across skills in the long-term learning process, we are inspired by the work of DKT-DSC~\cite{minn2018deep}. We reformulate the ability profile of a student and simplify it without sacrificing its originality and performance. It divides the student's interactions into multiple time intervals, then encodes student past performance for estimating their ability profile at the current time interval. The ability profile is encoded as a cluster ID and computed from the performance vector (with Equation~\ref{equ:seg}) of length equal to the number of skills, and updated after each time interval by using all previous attempts on each skill. The  success rates on each skill from past attempts data are transformed into a performance vector for clustering student $i$ at time interval $1\!\!:\!\!z$ as follows (for brevity we omit indexing all terms by $i$ in Equation \ref{equ:segcorr}):

\begin{equation}
R(x_{j})_{1:z}=\sum_{t=1}^{z}\frac{(x_{jt})}{|N_{jt}|},\label{equ:segcorr}
\end{equation}%
\begin{equation}
d_{1:z}^{i}=(R(x_{1})_{1:z},R(x_{2})_{1:z},...,R(x_{n})_{1:z}),\label{equ:seg}
\end{equation}

\noindent
where 
\begin{itemize}
	\item $x_{jt}$ is the outcome of attempt of skill~$x_{j}$ being correctly answered at time interval $t$; 1 denotes successful attempts and 0 denotes unsuccessful attempts;
	\item $|N_{jt}|$ is the total number of practice attempts of skill~$x_{j}$ up to time interval $z$;
	\item $n$ is the total number of skills;
        \item $R(x_{j})_{1:z}$ represents the ratios of skill~$x_{j}$ being correctly answered from time interval 1 to current time interval~$z$ by student~$i$. This is computed for all skills $(x_{1},x_{2},..,x_{n})$;
        \item $d_{1:z}^{i}$ represents a \textit{performance vector}  of student~$i$ on   all skills from time interval 1 until~$z$.
  
\end{itemize}

Each student has a different number of total time intervals in the lifetime of their interactions with the system.

If a student's time interval has no attempt in the interval ${0\!\!:\!\!z}$, we assign 0.5 to ratio $R(x_{j})_{1:z}$.

Therefore, data contains the encoded vector of the student's past performance and is accumulated and updated after each time interval. Time interval $z$  and student $i$ are ignored in the training process and only used in the clustering process later. Then the $k$-means algorithm is used to evaluate the temporal long-term learning ability of students in both training and testing at each time interval $z$, by measuring the Euclidean distance with centroids achieved after the training process as in  DKT-DSC~\cite{minn2018deep}.

After learning the centroids of all clusters, each student $i$ at each time interval $z$ is assigned to the nearest cluster $C_c$ by the following equation:
\begingroup\makeatletter\def\f@size{9}\check@mathfonts
\begin{equation}\label{equ:cluster}
\textbf{ability profile}(ab_z) = \argmin_{C} \sum^K_{c=1} \sum_{d^i_{1:z-1}\in C_c} ||d^i_{1:z-1}- \mu_c||^2
\end{equation}
\endgroup
\begin{figure}[t]
	\centering \includegraphics[width=1\linewidth, trim=0mm 10mm 0mm 10mm]{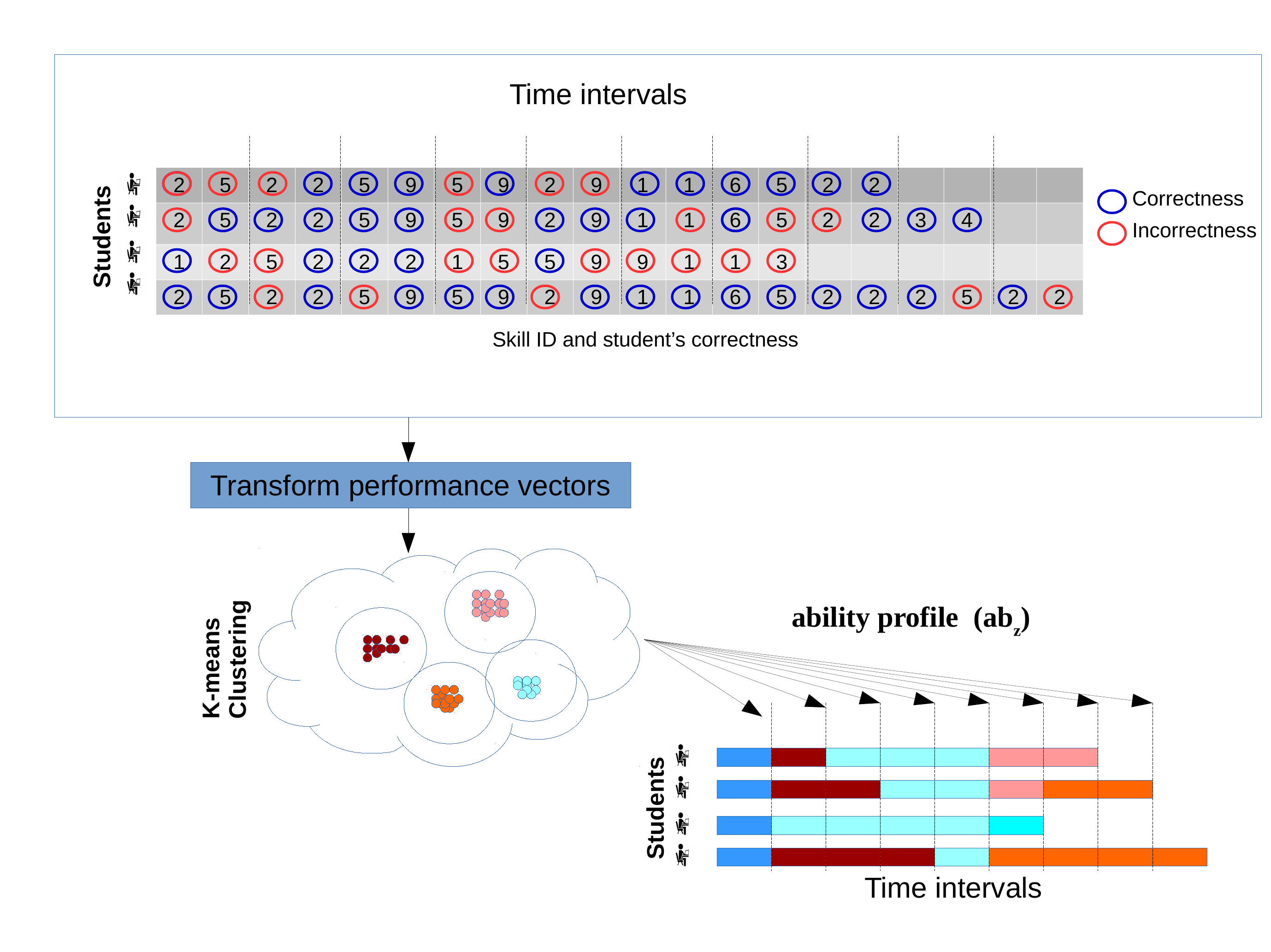} \caption{Detection of student's ability profile at each time interval.}
	\label{fig:clustering} 
\end{figure}
\noindent
where centroid $\mu_c$ is the mean of points for cluster $C_c$ , and performance vector $d^i_{1:z-1}$ is the average performance data of student $i$ from time interval $1$ to $z-1$.

Students are assigned to the nearest cluster and the label of this cluster $ab_z$ represents the temporal student learning ability at time interval $z$. Evaluation is started after the first 20 attempts and then every 20 attempts made by a student. For the first time interval, all students are assigned with initial ability profile 1.

By adding this cluster ID $ab_{z}$ (ability profile) of what group the student belongs to, we ensure that these high-level skill profiles are available to the model for making its predictions throughout the long-term interaction with the tutor.

\subsubsection{Estimating Problem Difficulty}\label{sec:difficulty}

The problem difficulty serves as a distinct feature for predicting predict student performance in previous studies ~\cite{minn2018improving,minn2019dynamic}. Note that, in this study, we assume each problem is associated with a single skill, but the difficulty is associated with problems, not with the skills themselves. The difficulty of a problem $p_j$ is determined on a scale of 1 to 10. $\textbf{Problem difficulty}(P_j)$ is calculated as:

\begin{equation} \label{equ:incorrratio} 
\textbf{difficulty level}(p_j) = 
\begin{cases}
\delta(p_{j}) & \text{if}\ \rvert N_j \lvert \geq 4\\
5 & \text{otherwise}
\end{cases}
\end{equation}
where:
\begin{equation} \label{equ:catdelta}
\delta(p_{j}) = \left\lfloor  { \frac{{ \sum_i^{\lvert N_j \rvert }}{O_i(p_{j})}}{\lvert N_j \rvert}} \cdot 10   \right\rfloor
\end{equation}
and where 
\begin{itemize}
	\item $p_j$ represents the $j^{th}$ problem
	\item $N_j$ is the set of students who attempted problem~$p_j$
	\item $O_i({p_{j}})$ the outcome of the first attempt from student~$i$ to problem~$p_j$, 1~if successful, 0~otherwise.

\end{itemize}

$\delta(p_{j})$ is a function that maps the \emph{average success rate} of problem~$p_j$ onto $10$ levels. Unseen problems, those that do not have any record, and problems with fewer than 4~students ($ \rvert N_j\lvert < 4$) in the dataset will have a difficulty of 5.



\subsection{Interpretable Student Performance Prediction}

\begin{figure}
	\centering \includegraphics[width=8cm, trim=0mm 30mm 0mm 30mm]{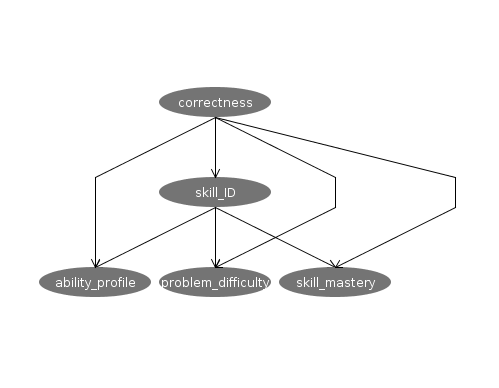}  \caption{Tree Structure of TAN.}
	\label{fig:TAN} 
\end{figure}

In order to get an interpretation with diagnostic and prognostic reasoning, we decide to choose the Bayes net paradigm for future development. So, our approach utilizes a Tree-Augmented Naive Bayes Method~\cite{friedman1997bayesian}. The TAN structure is a simple extension of the Naive Bayes network. Like Naive Bayes, the root node is the class node (correctness of the problem), causally connected to evidence nodes (skill ID, skill mastery, ability profile, and problem difficulty). Additionally, the TAN structure relaxes the assumption of independence between the evidence nodes~\cite{minn2014efficient}. It allows most evidence nodes to have another parent,
which can be a related evidence node. This model inherits the
directed acyclic graph structure~\cite{minn2016algorithm,sein2016accelerating} and produces
a tree that captures relationships among the evidence nodes.
The learning of this structure is not as computationally
expensive as a general Bayesian network and much more cost effective than building a neural network for knowledge tracing in DKT.
An example TAN structure is illustrated in Figure \ref{fig:TAN}.
The class node is the student's correctness hypothesis under consideration. The other nodes represent supporting evidence for the particular student's performance hypotheses at time $t$. Dependencies among the evidence nodes are captured as additional causal links in the TAN structure. Even though the direction of arrows represents the causal links between two nodes, information can flow in any direction based on the reasoning process~\cite{pearl2001bayesian}. 

The structure of TAN can be learned in different ways: 
\begin{itemize}
    \item a greedy search with the constraint that a node having more
than one parent from the evidence nodes is not allowed~\cite{cohen2004correlating};
    \item a Minimum Weighted Spanning Tree (MWST) approach that builds a minimum spanning tree to capture the dependencies among evidence nodes, and
then connects the class node to all of the evidence nodes~\cite{friedman1997bayesian}.
\end{itemize}

The MWST algorithm is applied in this research by using the data mining toolkit Weka~\cite{hall2009weka}. Learning the structure of TAN is done by only using training data.

Our model predicts whether a student will be able to answer the next problem $p_{t}$ based on their current knowledge state $\textbf{skill mastery}(s_t)$, learning transfer across skills $\textbf{ability profile}(ab_z)$ and difficulty level of problem occurring $\textbf{problem difficulty} (P_j)$. Instead of only feeding all student previous interactions $X=(x_1,x_2,..,x_t)$ to a neural network, we propose a novel model called IKT. 

IKT performs inference by using three meaningful extracted features $f_t$: skill mastery, ability profile, and problem difficulty as evidence at the current timestamp $t$.

\begin{equation}\label{equ:IKT1}
P(correctness_t=y|f_t) = \frac{P(y) P(f_t|y)}{\sum_{y'}P(y') P(f_t|y')}
\end{equation}

\begin{align*}\label{equ:IKT2}
\textnormal{where } P(f_t|y)  = & P(s_t|y) P(\textbf{ability profile}(ab_z)|y,s_t) \\ & P(\textbf{problem difficulty}(P_j)|y,s_t)  \\ & P(\textbf{skill mastery}(s_t)|y,s_t)
\end{align*}

Those are assessed $\textbf{skill mastery} (s_t)$ of student $i$ on skill $s$ at time $t$, the temporal $\textbf{ability profile}(ab_z)$ of student $i$ at current time interval~${z}$, and $\textbf{problem difficulty} (P_j)$ of problem $P_j$ at time $t$ (for brevity we omit indexing all terms by student $i$, $s_t$, $ab_z$ $P_j$ in Figure \ref{fig:TAN} and Equation \ref{equ:IKT1}).  The inference is estimated in the context of discretized values, i.e. conditional probability tables~\cite{hall2009weka}. It doesn't handle continuous variables. Discretization algorithm bin all features into sets for best discrimination among classes~\cite{mack2011using}.

The class node (correctness) represents the predicted probability that the student would answer the problem with the associated skill correctly. Thus the prediction of problem associated with skill $s_{t}$ can be retrieved from $correctness_t$ as described in Figure~\ref{fig:TAN}. We can achieve interpretation via the conditional probability tables of each node with their causal links. We can trace back the cause of failure in students' problem-solving by detecting whether there is a deficiency in practicing skills or the problem is too difficult for individual students with their evidence at each timestamp.

\section{Experiments}\label{sec:experimental-study}

We compare the next problem student performance prediction of our model with well-known KT models mentioned above:  BIRT~\cite{wilson2016back}, BKT~\cite{corbett1994knowledge}, PFA~\cite{pavlik2009performance}, DKT~\cite{piech2015deep}, DKT-DSC~\cite{minn2018deep} and DKVMN~\cite{zhang2017dynamic}. But we do not compare with other variants, because they are more or less similar and do not show significant performance differences. Table \ref{tab:comparison1} summarizes the characteristics of compared student models, in which each model has its significant characteristics.

\begin{table}[t]
	\caption{Comparison on characteristics of student models.}
    \setlength{\tabcolsep}{0.2em}
	\label{tab:comparison1} 
	\begin{centering}
		\begin{tabular}{|r|c|c|c|}
			\hline 
			
			Models & Ability profile & Problem information   & Sequential   \tabularnewline
			\hline

			IRT & No & Yes &  No   \tabularnewline
			PFA &  No & No &  No  \tabularnewline
			BKT &  No & No &  Yes  \tabularnewline
			DKT &  No & No & Yes \tabularnewline
		    DKT-DSC &  Yes & No & Yes \tabularnewline
		    DKVMN &  No & No & Yes \tabularnewline
		    AKT & No & Yes & Yes \tabularnewline
		    IKT & Yes & Yes & Yes \tabularnewline
			\hline 
		\end{tabular}
		\par\end{centering}
	
	\vspace{-2mm}
\end{table}

We implement all NN models with Tensorflow and all LSTM-based models share the same structure of fully-connected hidden nodes with an embedding size of 200. For speeding up the training process, mini-batch stochastic gradient descent is used to minimize the loss function. The batch size for our implementation is 32. We train the model with a learning rate of 0.01 and dropout is also applied for avoiding overfitting. We set the number of epochs to 100. All the models are trained and tested on the same sets of training and testing students. For BKT, we learn models for each skill and make predictions separately  and the results for each skill are averaged.

In this experiment, we assume 20 attempts made by a student as a time interval for that student for detecting the student's ability profile. The total number of temporal values for students' ability profiles used in our experiment is 8 (7 clusters and 1 for initial ability profile before evaluation in initial time interval for all students).  Five-fold cross-validation is used to make predictions on all datasets. Each fold involves splitting into 80\% training students and 20\% test students of each dataset. We conduct our analysis through Area Under Curve (AUC) as the majority class (correct) ranges 65-75 \% across all datasets, and Root Mean Squared Error (RMSE) as it is a variant of Brier score commonly encountered in the knowledge tracing community~\cite{gervet2020deep,bergner2021multidimensional}. For the input of DKVMN, values in both key and value memory are learned in the training process. For other models, one-hot encoded method is applied. The learned embeddings in value memory represent pre-trained weights which may contain the difficulty of each skill. For IKT, we learn the tree structure of TAN by using training data and making inference in testing data with WEKA~\cite{hall2009weka}.  

\subsection{Datasets}

\label{sec:dataset}

In order to validate the proposed model, we tested
it on three public datasets from two distinct tutoring scenarios in
which students interact with a computer-based learning system in
educational settings: 1) ASSISTments\footnote{\url{https://sites.google.com/site/assistmentsdata/}} is 
 an online tutoring system that was first created in 2004 which
engages middle and high-school students with scaffolded hints in their
math problem. If students working on ASSISTments answer a problem
correctly, they are given a new problem. If they answer it incorrectly,
they are provided with a small tutoring session where they must answer
a few questions that break the problem down into steps. Datasets are as follows: ASSISTments 2009-2010 (skill builder), ASSISTments 2012-2013~\cite{feng2009addressing}.
2) Cognitive Tutor. Algebra 2005-2006\footnote{\url{https://pslcdatashop.web.cmu.edu/KDDCup/downloads.jsp}} is a  development dataset released in KDD Cup 2010 competition from Carnegie Learning of PSLC DataShop.

\begin{table}
	\caption{Overview of datasets.}
	\label{tab:data} 
	\begin{centering}
		\begin{tabular}{|c|r|r|r|r|}
        	\hline
			\multirow{3}{*}{Dataset} & \multicolumn{4}{c|}{Number of} \tabularnewline
			\cline{2-5} 
			& \multicolumn{1}{c|}{Skills}  & \multicolumn{1}{c|}{Problems} & \multicolumn{1}{c|}{Students}& \multicolumn{1}{c|}{Records}  \tabularnewline
			\hline 
		    Algebra & 437 & 15663 & 574  & 808,775  \tabularnewline
			\cline{1-1} 
			ASS-09 & 123 & 13002  & 4,163  & 278,607   \tabularnewline
			ASS-12 & 198 & 41918 & 28,834  & 2,506,769  \tabularnewline
			\hline 
		\end{tabular}
		\par\end{centering}
	\vspace{-2mm}
\end{table}

 For all datasets, only the first correct attempts to original problems are considered in our experiment. We remove data with missing values for skills and problems with duplicate records. To the best of our knowledge, these are among the most well known publicly available knowledge tracing datasets.

\subsection{Results}

\begin{table}[t]
	\begin{center}
		\caption{AUC result for all tested datasets.}\label{tab:exp1}
		\begin{tabular}{|c|ccc|c|}
			\hline
			
			\multirow{3}{*}{Models} & \multicolumn{3}{c|}{Datasets}& \multicolumn{1}{c|}{} \\
			\cline{2-4}
			&  ASS-09& ASS-12 & Algebra  & Average      \\
			\hline
			BIRT & 0.750 & 0.744 & 0.812  & 0.768  \\  
			
			PFA &  0.701 & 0.672 &  0.754 & 0.709 \\
			\hline
			BKT & 0.651  & 0.623  & 0.642 & 0.638 \\  
			
			DKT & 0.721  & 0.713 & 0.784 & 0.739  \\
			
			DKT-DSC &  0.735 & 0.721 &  0.792 & 0.749\\
			
			DKVMN &  0.710 & 0.707 &  0.780 & 0.732\\
			
			AKT-R &\underline{0.767} & \textbf{0.777}  & \underline{0.845} & 0.796 \\
			\hline
			
			\textbf{IKT-3} & \textbf{0.797} & \underline{0.767}  & \textbf{0.851} & \textbf{0.805} \\
			\hline

		\end{tabular}
		
		\begin{tablenotes}
        \small 
        \item{Best scores are in bold, second best scores are underlined.}
        \end{tablenotes}
        
	\end{center}	
\end{table}

The TAN structure results in a student model with better explanation with  causal relations and higher predictive performance. The results in Tables \ref{tab:exp1} and \ref{tab:exp2} demonstrate that IKT outperforms significantly the well-known KT models in all tested datasets. IKT-3 has superior performance than any other models tested in our experiments. When we compare IKT-3 with our second best performer AKT-R (scores are underlined in table \ref{tab:exp1}), the improvement is ranging from 0.71\% to 3.91\% in Algebra and  ASS-09 datasets in terms of AUC. When we compare in terms of RMSE, it shows an improvement up to 2.84\% over the second best performer AKT-R on the ASS-09 dataset. So IKT shows better performance than any other method in both AUC and RMSE (except AKT-R on the ASS-12 dataset).

\begin{table}[t]
	\begin{center}
		\caption{RMSE result for all tested datasets.}\label{tab:exp2}
		\begin{tabular}{|c|ccc|c|}
			\hline
			
			\multirow{3}{*}{Models} & \multicolumn{3}{c|}{Datasets} & \multicolumn{1}{c|}{} \\
			\cline{2-4}
			&  ASS-09 & ASS-12 & Algebra & Average        \\
			\hline
			BIRT  & 0.440  & 0.441  & 0.374 &  0.418   \\  
			
			PFA & 0.454 & 0.440 & 0.391 & 0.428   \\
			\hline
			BKT &  0.471 & 0.510  & 0.440 & 0.473\\  
			
			DKT & 0.450  & 0.430  & 0.380 & 0.420  \\
			
			DKT-DSC &  0.434 & 0.427 &  0.373 & 0.411\\
			
			DKVMN & 0.451 & 0.430 & 0.380 & 0.42 \\
			
			AKT-R & \underline{0.423} & \textbf{0.409} &  \textbf{0.354} & \underline{0.395} \\

			\hline
			
			\textbf{IKT-3} & \textbf{0.411}  & \underline{0.413}   & \textbf{0.354} & 
			\textbf{0.392}  \\
			
			\hline

		\end{tabular}  
		
		\begin{tablenotes}
        \small 
        \item{Best scores are in bold, second best scores are underlined.}
		\end{tablenotes}
	\end{center}	
\end{table}

\subsection{Ablation studies}\label{sec:comp-across-diff}

The results so far suggest there may be a different impact of each factor on the predictive performance of KT models, and in particular the impact of item difficulty. This question is further analyzed in this section. 

We compare our IKT model through an ablation study with the following different features:
\begin{itemize}
	\item IKT-1: skill ID, skill mastery.
	\item IKT-2: the features of IKT-1 + ability profile.
	\item IKT-3: the features of IKT-2 + problem difficulty.

\end{itemize}



This study helps us understand the contribution of each feature in student performance prediction. IKT-1 takes only skill ID and skill mastery into account for student performance prediction which achieves higher performance than original BKT and has a bit lower performance than DKT (where DKT takes only binary values of student previous interaction). When IKT-2 takes skill ID, skill mastery and ability profile of a student, it shows similar (or a bit lower) performance as DKT and higher performance than BKT, PFA on two Assistments datasets. Our proposed model  IKT-3 shows better performance than any KT methods compared in this experiment.

\begin{table}[t]
	\begin{center}
		\caption{AUC and RMSE result for ablation study.}\label{tab:exp3}
		\begin{tabular}{|c|ccc|}
			\hline
			
			\multirow{3}{*}{Models} & \multicolumn{3}{c|}{AUC} \\
			\cline{2-4}
			&  ASS-09 & ASS-12 & Algebra        \\
			\hline
			
			IKT-1 & 0.705  & 0.690   & 0.731 \\  
			
			IKT-2 & 0.715   & 0.696   & 0.734    \\
			
			\textbf{IKT-3} & \textbf{0.797} & \textbf{0.767}  & \textbf{0.846} \\
			
            \hline
			
			\multirow{3}{*}{Models} & \multicolumn{3}{c|}{RMSE} \\
			\cline{2-4}
			&  ASS-09 & ASS-12 & Algebra        \\
			\hline
		
			IKT-1 &  0.443  & 0.437   & 0.395  \\  
			
			IKT-2 &  0.441   & 0.435   & 0.394    \\
			
			\textbf{IKT-3} & \textbf{0.411}  & \textbf{0.413}   & \textbf{0.354}  \\

			\hline

		\end{tabular}  
	\end{center}	
\end{table}

Table \ref{tab:exp3} reports the results. Ability profile results in a mild improvement in AUC smaller than 1.4\% between IKT-1 to IKT-2. Results from each model with different combination of features show us which features provide more information in student performance prediction among various datasets. We can also see that the problem difficulty factor is the most influential factor to student performance prediction. When we apply problem difficulty to IKT-3, it increases around  11.2\% to 15.7\% in AUC (e.g. 0.731 AUC in IKT-1 to 0.846 IKT-3 in Algebra) respectively. It also explains why both BIRT and AKT-R have a better performance in student performance prediction among other models (see Tables \ref{tab:exp1}, \ref{tab:exp2} and  \ref{tab:exp3}).

\section{Conclusion}\label{sec:conclusion}

We extract three meaningful latent features from students' behavioral data by using data mining techniques in feature engineering. The TAN structure with causal relations based on these features results in a KT model with better performance for student performance prediction, that does not require a huge number of parameters nor a complex structure. It saves huge computational resources compared to deep learning models and provides a causal explanation for better understanding with meaningful features. 

We proposed a causal probabilistic student model called IKT with three extracted meaningful features: student's skill mastery (probability of learning a skill),  ability profile (learning transfer of a student), and problem difficulty to predict student performance. Unlike deep learning-based KT models, which only take students past interaction and learn all information in a hidden state with a huge amount of parameters and complex relations in structure, IKT can predict student performance by utilizing extracted features from skills, problems and students and gives us meaningful causal explanations. Experiments with three public datasets show that the proposed model outperforms well-known KT models (including deep learning-based models) and requires less computational power than deep learning-based models. Feature engineering helps us capture more variance from the data and Tree-Augmented Naive Bayes plays a critical role in providing causal explanations. It leads to more accurate and personalized performance predictions.

Although we focused only on prediction performance, we are also interested in studying the causal effect of the meaningful features, aiming to serve adaptive instructions in personalized learning systems. Meanwhile, a trade-off between student prediction performance and causal explanation should be found. More experiments
will be designed and conducted to understand more about knowledge acquisition and learning behaviors to optimize human learning in future educational environments.

\bibliography{aaai22}

\end{document}